# An Ontology Oriented Architecture for Context Aware Services Adaptation

Hatim Guermah[1], Tarik Fissaa[1], Hatim Hafiddi [1,2], Mahmoud Nassar[1] and Abdelaziz Kriouile[1]

[1] IMS Team, SIME Lab, ENSIAS, Mohammed V Souissi University,
Rabat, Morocco

[2] ISL Team, STRS Lab, INPT,
Rabat, Morocco

**Abstract**
In the field of ubiquitous computing, a class of applications called context-aware services attracted great interest especially since the emergence of wireless technologies and mobile devices. Context-aware application can dynamically capture a range of information from its environment and this information represents a context, the application adapts its execution according to this context. An important challenge in ubiquitous computing is dealing with context. Ontologies presents the most promising instrument for context modeling and managing due to their high and formal expressiveness and the possibilities for applying ontology reasoning techniques. In this paper, we present an ontology based approach for the development of context aware services.

## 1. Introduction

In ubiquitous computing system, a set of intelligent computing devices communicate and collaborate by perceiving the global context and responding proactively (without explicit user intervention) to provide adapted services to the user and other applications. Thus, in such system, the equipment must be context-aware. Therefore, the context is a key concept in such systems and thus requires a good understanding and use. Despite the large number of proposed definitions, This term is still vague and general.

Context aware applications have emerged in the ambition to provide access to the system any time, anywhere and with a content corresponding to the user's context. Such environments have as objective to optimize user interaction with embedded devices for example by allowing users access to all available information and adapting them to the actual conditions (location, whether.. ). This requires applications to dynamically adapt their behavior according to environmental characteristics.

Many approaches have been proposed to deal with Context Aware Services (CAS), but most only offer weak support for knowledge sharing and context reasoning. Ontologies have proved to be the most suitable model for representing and reasoning on context information for the following reasons [1]:

i. ontologies enable knowledge sharing in open, dynamic systems;
ii. ontologies with well-defined declarative semantics allow efficient reasoning on context information;
iii. ontologies enable service interoperability.

For the above reasons, CAS development can benefit from Ontologies, and Semantic Web Services. In this work, we aim to propose in first stage an overview of an architecture for the development of context aware services based on ontologies, we present our context ontology and in the second stage we discuss about context reasoning, at the end we present our contribution in semantic context-aware service by extending WSMO [2].

The remainder of this paper is structured as follows. Next we present a background about context awareness. Section 3 present some related works. In section 4 we present a motivation scenario that concerns context-aware E-health system, and highlight the context-awareness challenges. Then in section 5 we describe the proposed architecture and highlight its different component and we present our proposal for service adaptation by extending WSMO. The last section gives rise to some concluding remarks and plans for future works.

## 2. Background

### 2.1 Context

Several works have attempted to formalize the meaning of context in the computing area. However, a universally accepted definition is yet to be agreed. Researchers enumerated certain context types, they considered important and relevant. Schilit [3] defined context to be: location, identities of nearby people, objects and changes to these objects.

Other definitions have been proposed, Brezillon [4] define context as an information that characterizes the interactions between humans, applications and the environment. Dey [5] discuss that the important aspects of context cannot be enumerated, as they differ from situation to situation and depend on the purpose of the application, furthermore they formally defined context as:

> "... any information that can be used to characterize the situation of an entity. An entity is a person, place or object that is considered relevant to the interaction between a user and an application, including the user and applications themselves."

In our work we will adopt this definition because it remains the most generic.

## 2.2 Context awareness

In general, context awareness refers to the ability of an application to discover and take advantage of contextual information, such as user location and nearby devices. Shilit [3] is the first to introduce the concept of context-awareness and have defined it as the ability of an application to discover and react to changes according to the users environment. Brown [6] defines it as applications whose behavior can change depending on the user's context. Dey [5] considers an application as context aware if it uses contextual information to provide relevant information and services to the user, where relevance depends on the user's task.

In order to exploit the context in a relevant way and to have a reliable approach for the development of context-aware services, several challenges arose:

- Context capture: recovering properties that characterize the user context.
- Context representation: proposing representation that provides high-level abstractions.
- Context interpretation and reasoning: reasoning about Context consists in deriving Context from the existing one at a high semantic level.
- Service adaptation : through generated situations, services must be invoked adapted
- Context management: managing context consist of dealing with non-functional aspects
- Context reuse: taking advantage of contextual properties by demanding those validity has expired.

## 3. Related Work

## 3.1 Context modeling approaches:

In this section we will discuss the most relevant context modeling approaches.

## 3.1.1 Key-value models:

The model of key-value pairs is the most simple data structure for modeling contextual information. Already Schilit [3] used key-value pairs to model the context by providing the value of a context information (e.g. location information) to an application as an environment variable. In particular, key-value pairs are easy to manage, but lack capabilities for sophisticated structuring for enabling efficient context retrieval algorithms.

## 3.1.2 Markup Scheme Models:

Common to all markup scheme modeling approaches is a hierarchical data structure consisting of markup tags with attributes and content. Some of them are defined as extension to the Composite Capabilities/Preferences Profile (CC/PP) [7] and User Agent Profile (UAProf) [8] standards, which have the expressiveness reachable by RDF/S and a XML serialization.

## 3.1.3 Graphical Models:

A very well-known general purpose modeling instrument is the Unified Modeling Language (UML) which has a strong graphical component. Due to its generic structure, UML is also appropriate to model the context. This is shown for instance by Bauer in [9], where contextual aspects relevant to air traffic management are modeled as UML extensions. Another example is the nicely designed graphics oriented context model introduced in [10] by Henricksen, which is a context extension to the Object-Role Modeling (ORM) approach according some contextual classification and description properties.

## 3.1.4 Object Oriented Models:

Common to object oriented context modeling approaches is the intention to employ the main benefits of any object oriented approach - namely encapsulation and reusability – to cover parts of the problems arising from the dynamics of the context in ubiquitous environments

## 3.1.5 Logic Based Models:

A logic defines the conditions on which a concluding expression or fact may be derived (a process known as reasoning) from a set of other expressions or facts. In a logic based context model, the context is consequently defined as facts, expressions and rules. Usually contextual information is added to, updated in and deleted from a logic based system in terms of facts or inferred from the rules in the system respectively. Common to all logic based models is a high degree of formality.

## 3.1.6 Ontology Based Models:

As the context may be considered as specific kind of knowledge, it can be modeled as ontology. Ontologies are a very promising instrument for modeling contextual information due to their high and formal expressiveness and the possibilities for applying ontology reasoning techniques. There are several ontology based approaches among them CONON (stands for CONtext Ontology)[11], CoBrA-ONT (Context Broker Architecture ONTology)[12].

## 3.2 Context aware engineering approaches:

Kapitsaki [13] present a survey that tries to distinguish different solutions for context aware engineering proposed by different researchers and developers. The approaches can be divided in the following categories:
- Middleware solutions and dedicated service platforms.
- Use of ontologies.
- Rule-based reasoning.
- Source code level programming/language extensions.
- Model-driven approaches.
- Message interception

There are, however, generic cases where more than one paradigm or pattern are used together in the same approach. Hence, the above approaches are not completely disjoint and a potential developer may opt for a combination of several techniques

### 3.2.1 Middleware solutions:

The development and execution of context-aware applications can be made through middlewares. Their role is to facilitate context-awareness by integrating into their architectures a context management layer and mechanisms for service adaptation.

Kjær [14] provides a survey of a chosen set of context-aware middleware systems, categorises, their properties and use according to a taxonomy. The categorization is based on surveying existing context-aware middleware. The surveyed systems are very different and some are pure middleware while others are more complete infrastructural systems offering services for managing entire physical environments.

The major categories of the taxonomy are:
- Environment: A middleware system makes explicit or implicit assumptions about the environment it is to be used in.
- Migration: Some of the systems merely provides mechanisms for migrating running code when the application decides, possibly based on context, while other systems migrate entities automatically based on context.
- Storage: Some systems provide a context-aware data store which order data based on context information, allowing it to be retrieved based on certain context-parameters
- Quality: measure of how well a service can be performed or how good data is.
- Reflective mechanisms: mostly known in programming languages, but some middleware systems offer reflection of different parts of the systems.
- Composition: doing component composition based on contextual events.
- Adaptation: When context-information is available, systems can adapt to changes in the context.

The context toolkit [15] is a generic middleware that enables the rapid prototyping of context-aware applications. It provides a set of abstract components (Widgets, Interpreters and aggregators) that can capture and interpret the context to make it exploitable representation to applications. However, it does not provide mechanisms for context reasoning, and mechanisms that can be used by applications to respond to various context changes.

Another example of middleware, named "Pace", proposed by Henricksen [16] provides a set of tools and components that assist context-aware application developers in their activities. It is based on the following components:
- Context Management Component: provides mechanisms for the interpretation and storage of context information. For context modeling, middleware Pace uses CML language (Context Modeling Language).
- Preference Management Component: Management of user preferences.
- Messaging framework: aims to facilitate communication between various components of this middleware and context-aware applications.
- Schema compiler toolset: provides a set of tools for code generation.

### 3.2.2 Source code level programming or language extensions:

This type of approach aims to enrich the logic with fragments of code responsible for the execution of the context and providing services with adaptive behavior.

Hirschfeld [17] has dealt with this type of approach by working on several types of language. Indeed, he proposed an approach based on two paradigms:
- Context-oriented Programming (COP): explicitly addresses the context, and provides mechanisms to dynamically adapt the behavior in response to changing circumstances, even after the deployment of the system at runtime.
- Multidimensional approach: The term "layer" refers to a set of partial definitions of class and method with a specific behavior.

The system relies on a separate infrastructure to capture the context information from sensors and taking them to the place where the code is executed. In order to use these layers, code snippets are added at the layer activation time (construction "with") or deactivation (construction "without"). In this way, the behavior of the program can be changed dynamically during the execution in a context-aware manner.

Tanter [18] proposes a new paradigm, Context-Aware Aspect, able to adapt behavior according to the current context. Thus,

the aspects design run when specific settings are verified, and requires the ability to define the context with constructing cuts points. Indeed, the context is represented by a set of objects.
Such approaches generally focus on adaptation and do not provide mechanisms for capturing, representing or reasoning on context nor the reuse of contextual information.

### 3.2.3 Model-driven approaches:

Several approaches focus on the meta-modeling to meet the requirements of context-aware systems. The Model-Driven Engineering (MDE) paradigm aims at the definition of domain-specific modeling languages, transformations between meta-models for the production of executable code. In the development process the focus is given on the platform-independent modeling of the application that drives the transformation to the application code.
ContextUML [19] is one of the first approaches to modeling the interaction between context and Web services. ContextUML is a UML meta-model, which extends the existing UML syntax by introducing appropriate objects to enable the creation of context-aware service.

Thus ContextUML can be used to define:
- The allocation of associated parameters with contextual operations
- Management context awareness for services.

Ayed [20] specifies a UML profile approach to designing context-aware applications, platform independent. It offers a design process context models, but does not specify the mechanism of adaptation.
Simons [21] proposes CMP is a context modeling that uses stereotypes to extend the UML class diagram to model context. This model formalizes the meta information of context (ie source and validity of context information) and reflect privacy restrictions. The profile provides several well-formed rules for context models supporting the development of context-aware applications through a graphical modeling language.
Such approaches mainly offer design process for context models, but does not specify the mechanism of adaptation to meet the context changes.

### 3.2.4 Ontology based approach

Ontologies open a new area to deal with context-aware services. As the context may be considered as a kind of specific knowledge, it can be modeled as an ontology. Indeed, the use of ontologies allows not only context modeling, but also to reason, based on an inference engine, on the collected data.
SOCAM [22] (Service Oriented Context-Aware Middleware) is a middleware for context-awareness that provides an infrastructure for creating context-aware applications. SOCAM has also implemented a context reasoning engine that reasons over the knowledge base. SOCAM discharges developers of context-aware services from the task of context management by providing a model to describe the context. However, there is no way for developers to choose the sensors used to collect context. The adaptation is not taken into account and left to the developers who should implement the adaptations of each service.
SOCAM use CONON (stands for CONtext Ontology) an ontology divided into two levels. The first one is called upper ontology and describes general concepts which are common to all context-aware applications, while the second one is extensible to add domain specific ontologies, for reasoning and they use logic based reasoning to derive high-level context.

### 3.2.5 Rule based reasoning:

This type of approach aims to obtain new results from a combination of rules and a set of activation conditions of these rules by reasoning on contextual parameters.
The context handling platform [23] is based on the Model Event-Control-Action (ECA). This architectural model is designed to provide a structural diagram to enable the coordination, cooperation and configuration of distributed functionality in a context aware platform. The ECA pattern divides the tasks of gathering and processing context information, from tasks of triggering actions in response to context changes. These separate tasks are realized under the control of an application behavior description, in which reactivation of context-aware application behaviors are described in terms of ECA rules, which have the form if-then-else. The action part of the rule is composed by one or more actions that are triggered whenever the condition part is satisfied. This approach proposes a specific language called ECA-DL in order to specify the rules of the ECA.
This type of approach is very interesting because it allows to obtain high level context, however this technique don't deal with context modeling and the service adaptation

### 3.2.6 Message interception:

Message interception allows service adaptation according to the context by intercepting input/output messages and modifying them without affecting basic services.
Kapitsaki [24] present an architecture, based on Axis2 web service framework, that intercepts the service requests and responses, retrieves the context information related to these messages and returns a modified message reflecting the context adaptation. The message modification is carried out through a number of plugins. The context plugins communicate with the context sources that have direct access to the context information and alter the input and output messages to reflect the current user task and its environment. So, the context adaptation is kept independent from the specific service implementation and clearly separated from the service logic.
This technique is very interesting because in one part, it is domain independent and in the other part it allows service adaptation according to contextual information. However, it's still limited because the lack of context modeling and reasoning strategy.

### 3.2.7 Hybrid approaches:

The combination of various approaches allows dealing with ubiquitous computing requirements by treating the context from the capture phase to the adaptation phase as well as other non-functional aspect.

Hafiddi [25] define metamodels for modeling context-aware applications by planning several model views that model system context sensitivity. The authors propose an architecture for Context-Awareness of Services (CAS), this architecture relies on a set of context-awareness and CAS specifications and metamodels to enhance a core service, in service oriented systems, to be context aware. Moreover, the semantic dimension remains ignored in this proposed architecture.

## 4. E-Health Motivating Scenario

With recent progress in the field of software engineering and the emergence of new ubiquitous mobile devices, the importance of context aware services raises. One of their application, is personalized search services offered by the health sector.

Let's imagine Mr John needs information about the nearest pharmacy to his position or hospital that provides medical servicse and the physician in charge of the service and its availability. He needs to connect via his mobile device (smartphone, tablet…etc) to an e-health system in order to obtain the appropriate medicals information (available pharmacies for example). the proposed solution must be able to use the information related to the user's context (location, device properties, profile…etc) to present an adequate list that meets the user's needs and takes into account the user's device and the possibility of transition to a reduce view to avoid blocking.

For the development of such E-health system, we have to choose the necessary context information and the source from which we can extract it. Next step is context modeling using ontologies to provide formal semantics to context. Then, it is important to reason about context at a semantic level to interpret this information and obtain interesting knowledge. According to this knowledge, the system adapts its behavior to meet the user's needs within ever-changing context.

## 5. Architecture overview

This architecture (Figure 1), relies on, a context ontology divided on two parts an upper ontology and a domain specific ontology that extends the first one, a processing and reasoning engine which defines the situations and the necessary adaptations, and a tool for service adaptations based on WSMO, the details of our architecture components:

- User Device: It is a smart device which is equipped with the context-aware sensors relied to it. It allows to enter user's request and acquire via sensors the necessary contextual information.
- Context Retrieval and Assimilation: Capture contextual information, inserted, captured or stocked, and manages their heterogeneity.
- Context Processing & Reasoning: reasoning and deducing new situations from OWL representations of context and inference rules.
- Service Adaptation: from situations derived by Context Processing & reasoning, this tool allows us to adapt services, to context at runtime based on WSMO services.

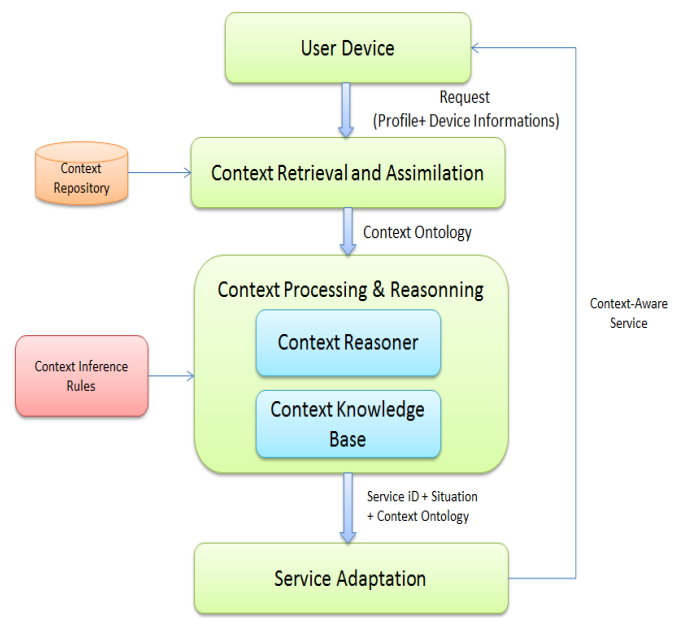

Figure 1. Overview of the proposed architecture

### 5.1 User Device

Modern mobile devices are graphically powerful devices with multiple sensors including GPS, accelerometers…etc that facilitate the creation of rich and engaging interaction design and user experiences by sensing context information around the user.

### 5.2 Context Retrieval and Assimilation

The notion of ontology, which is often used by artificial intelligence practitioners for knowledge representation, has emerged as a new approach for context modeling. It models the context at a higher semantic level to establish a common understanding of terms and meanings for sharing context, reasoning and reuse in ubiquitous Environments. Indeed, the aim is to convert the raw contextual data to a level where the semantic context can be shared and supplied to the context aware services.

Thus, we choose OWL as an ontology description language designed for publishing and sharing ontologies on and allowing a rich knowledge representation(based on properties, classes with identity, equivalence, opposite, cardinality, symmetry, transitivity…etc) and reasoning on this knowledge based on a formal axiomatic.

Thus, the proposed context ontology (Figure 2) is characterized by a two-level hierarchy. The first one is general and domain independent. It brings together the contextual properties to general categories of OWL classes independently of the domain. In our case, we specify five contextual classes:

- User : information about user's profile (name, age, etc....) and his preferences (language ..etc).
- Service : Describes the characteristics of the services requested by the user.
- Activity : describes the type of information requested by the user.
- Device : describes the user's device and its properties
- Environment : describes the environment-related properties.

The second level is domain-dependent and specific to some kind of applications. For example the class pharmacy or hospital is linked to the e-health domain.

To link between different classes in our ontology, we use several relations Subclass, part of and ComplementOf. Also, each class contains multiple properties, Object properties and Datatype properties, used to describe the characteristics and attributes of the concept and different types of restrictions for each property described by values or cardinality constraints.

Thus we get the ontology described in the Figure 2, it will be an input for the Context Processing and Reasoning component in order to infer high level context from low level one.

## 5.3 Context Processing and Reasoning:

The formalism of choice in ontology-based models of context information is typically OWL-DL or some of its variations, since it is becoming a defacto-standard in various application domains, and it is supported by a number of reasoning engines. By means of OWL-DL it is possible to model a particular domain by defining classes, individuals, characteristics of individuals (data type properties), and relations between individuals (object properties). Ontology based context reasoning is used in context approaches for deriving new context information on the basis of both OWL defined concepts and user defined rules.

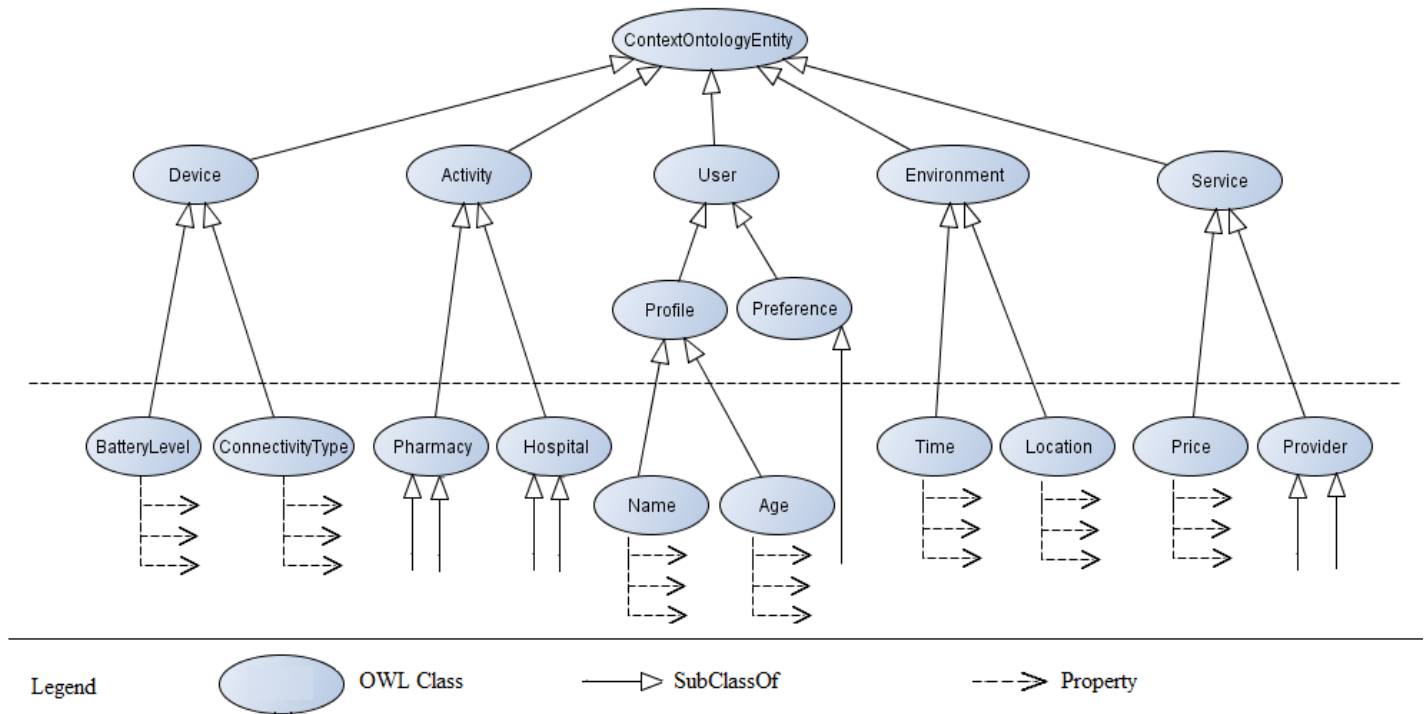

Figure 2. Context Ontology

Reasoning about Context consists in analyzing the repository of Context in order to derive context from the existing one at semantic level. For instance, given the current position of a user and the different mobility relationships we can derive all the locations to which the user can pass from its current position. In order to define these derivations, we use rules.
There are several reasoners or inference engine that allows to infer knowledge from OWL.

The context reasoning engine is composed of :
- Reasoner: allows to infer new situations from relevant contextual properties based on defined semantic relations and inference rules.
- Adaptation inference Rule : contains a set of situations
- Knowledge base: information repository for context that provides a means for information to be collected, organized, shared, searched and utilized.

Thus, for our scenario, the user has a mobile device (smartphone or tablet) able to collect contextual information and to send user's requests. To be used in health care services, this low-level context should be converted to high-level context according to the context model and inference rules (see Table I).

| Low level context | High level context | Rules |
|---|---|---|
| Location Time Pharmacy Medication | Nearest Pharmacy | If (PharmacyLocation=CurrentLocation) ^ (PharmacyOpened=False) ^ (MedicationDisponibility=False) |
| | Open Pharmacy | If (PharmacyLocation=CurrentLocation) ^ (Time=CurrentTime) ^ (PharmacyOpen=True) ^ (MedicationDisponibility=False) |
| | Medication Dispo InPharmacy | If (PharmacyLocation=CurrentLocation) ^ (Time=CurrentTime) ^ (PharmacyOpen= True) ^ (MedicationDisponibility=True) |

TABLE I : Example of Rules For Generation High level context

For instance, let take as an example the case of Pharmacy search. Many contextual information can affect the search result. Thus, the user can search the closest pharmacies or more specifically those that are currently open or more specifically that contains a particular medication.

## 5.4 Services Adaptation

Several specification and frameworks are interested in integrating ontology in the different development stages. WSDL-S (Semantic Web Services Description Language) [26], OWL-S (Semantic Web Ontology Language) [27] and WSMO (Web Service Modelling Ontology) [2] are the most known solution.

Based on DAML-S, OWL-S is a general ontology for building semantic Web services. It will enable users and software agents to automatically discover, invoke, compose, and monitor Web resources offering services, under specified constraints [26]. The OWL-S ontology defines three parts: the service profile, the process model and the grounding:

- The service profile: is used to expose services to client,
- The process model: describe the structure of the service like, inputs, outputs, pre-conditions and results of the service execution.
- The grounding: is about details communication such as, communication protocols, message formats, and port numbers.

WSDL standard, providers functional description of Web services and ignore semantic. WSDL-S is an incremental extension of WSDL in order to include semantic information. WSDL-S proposes a set of attributes and elements:
- Model Reference: an attribute that link a WSDL entity and a concept in some semantic model.
- Schema Mapping: an attribute that map schema elements of a Web services and their corresponding semantic model concepts
- Precondition: an element that represent a set of semantic statements that must be satisfied before an operation can be invoked.
- Effect: an element that informs us about expecting changes to occur upon invocation of a service.
- Category: an element that consist of service categorization information that could be used when publishing a service in a Web Services registry such as UDDI.

Web Service Modeling Ontology (WSMO) provides a conceptual framework and a formal language for semantically describing all relevant aspects of Web services in order to facilitate the automation of discovering, combining and invoking electronic services over the Web [2]. WSMO define four main elements:
- Web Services:  This is an equivalent of WSDL file. It describes the functional and behavioral aspects of a Web services.
- Goals:  The reason why the client is consulting a Web Service.
- Ontologies: A formal Semantic description of the information used by all other components.

- Mediators: Provides interoperability between different components with different ontologies.

In this paper, we chose WSMO because it allows, firstly, to facilitate the selection and sequencing of services, and secondly to model metadata produced and ensure interoperability between heterogeneous systems.

Thus we propose an extension of WSMO based on the situations generated by the inference engine to detect the appropriate services to an active situation and choose the most appropriate service.

Figure 3 illustrate our Semantic Context Aware Service (SCAS). The SCAS is based on the following specifications:

- The SCAS has a CAOntology, CAWebService, CAGoal and CAMediators.
- The CAGoal is related to a contextual Situation.
- A situation contains SimpleSituation and CompositeSituation.
- CompositeSituation is composed of other Situation
- Each SimpleSituation is related to contextual parameters: ContextProperty
- Each ContextProperty contains Contextual Attributes : Attribute.
- Each CAWebService contains an Adaptation.
- For a given CAWebService, a set of AdaptationRule is associated;
- An AdaptationRule can involve the execution of an ordered set of Adaptations.
- An Adaptation is related to a relevant Situation.

For instance (see Figure 4), in the E-health motivating scenario (c.f. Sect.4), the service SearchingPharmacy can provoke adaptation by FindingNearestAdaptation (i.e. Adaptation) whenever the situation NearestPharmacy is active. This situation requires three low level context Properties such as PreferenceLanguage, Time and Location.

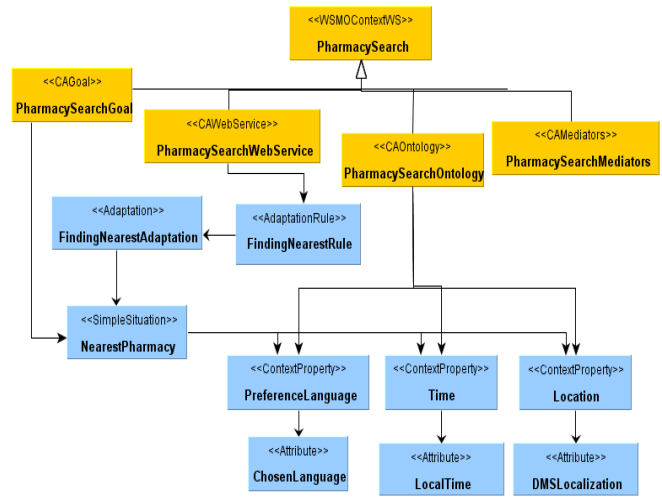

Figure 4. Illustration of Semantic Context-Aware Service

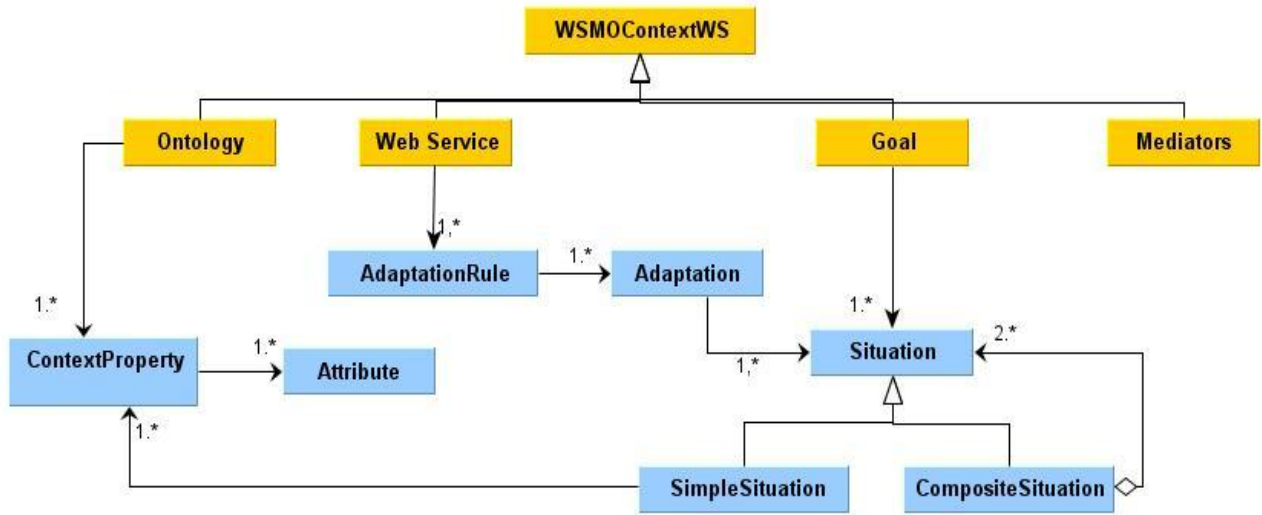

Figure 3. Semantic Context Aware Serivces by extending WSMO

## 5.5 Adaptation Tool

To build context aware services, we need to define mechanisms for the adaptation of their behaviour according to the current context situation. Such mechanisms will favourite loosely coupling between the core services and its adaptations seen as transversal preoccupations. The adaptations are eventually conditioned by the existence of relevant situations to the current context.

To meet these requirements, and inspired by the approach proposed in [25] which based on the Separation of Concerns and the Aspect Paradigm concepts considers the Adaptation as an aspect, we propose an evolved adaptation tool. Based on the situations generated by the inference engine and the context ontology, this tool allows to select which services to invoke, and adapt them.

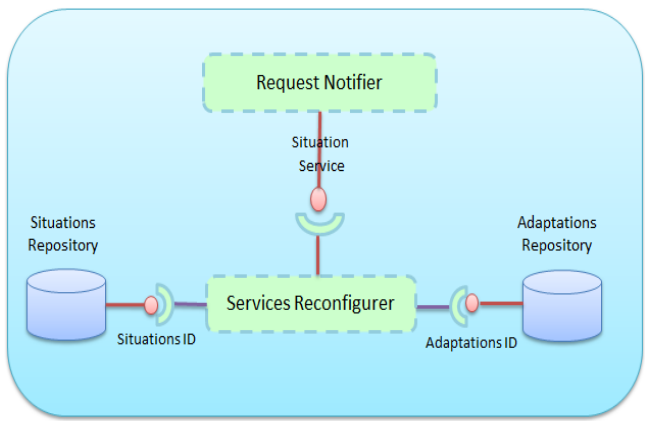

Figure 5: Adaptation Tool

Figure 5 illustrates the mechanism behind the Adaptation tool. The Request Notifier notifies, in a synchronous or asynchronous mode, the Service Identifier with the active situation and Id of Service in order to adapt this service to the given situation.

The interpretation mechanism operated by Reconfigure Service, recover situations and weave the necessary adaptation aspects, following a set of adaptation rules, and user context in services to produce a semantic contextual service.

## 6. Conclusion:

Context aware services are expected to play key roles in our society. It is an emerging paradigm for delivering services according to the user's context. A key enabler for those services is various types of contexts and reasoning with the contexts. Especially with the advent of sensor technology and availability in mobile devices, contexts become a key source of information from which situations can be inferred using reasoning techniques.

In this paper we presented a semantic approach for context aware services. First we defined an ontology to model context information, reasoning about this context to infer new situations and then performing service adaptation based on WSMO semantic services.

Currently we are implementing a framework to deal with context aware services. And we project to enhance this approach with new reasoning and adaptations mechanisms.

## 7. References:

[1] H. Chen, F. Perich, T. Finin, and A. Joshi, "Soupa: Standard ontology for ubiquitous and pervasive applications," in Mobile and Ubiquitous Systems: Networking and Services, 2004. MOBIQUITOUS 2004. The First Annual International Conference on. IEEE, 2004, pp. 258–267.

[2] Roman, D., Keller, U., Lausen, H.(eds.): Web Service Modeling Ontology (WSMO), available at http://www.wsmo.org/2004/d2/v01/index.html.

[3] B. Schilit, N. Adams, and R. Want, "Context-aware computing applications," in Mobile Computing Systems and Applications, 1994. WMCSA
1994. First Workshop on. IEEE, 1994, pp. 85–90.

[4] P. Brézillon, "Focusing on context in human-centered computing," Intelligent Systems, IEEE, vol. 18, no. 3, pp. 62–66, 2003.

[5] A. Dey and G. Abowd, "Towards a better understanding of context and
context-awareness," in CHI 2000 Workshop on The What, Who, Where, When, and How of Context-Awareness, 2000.

[6] P. Brown, J. Bovey, and X. Chen, "Context-aware applications: from the laboratory to the marketplace," Personal Communications, IEEE, vol. 4, no. 5, pp. 58–64, 1997.

[7] M. Nilsson, J. Hjelm, and H. Ohto, "Composite capabilities/preference profiles: Requirements and architecture," W3C Working Draft, vol. 21, 2000.

[8] WAPFORUM. User Agent Profile (UAProf). http://www .wapforum.org

[9] BAUER, J . Identification and Modeling of Contexts for Different Information Scenarios in Air Traffic, Mar. 2003. Diplomarbeit.

[10] Henricksen, K., Indulska, J., and Rakotonirainy, A. Generating Context Management Infrastructure from High-Level Context Models. In Industrial Track Proceedings of the 4th International Conference on Mobile Data Management (MDM2003) (Melbourne/Australia, January 2003), pp. 1–6

[11] X. Wang, D. Zhang, T. Gu, and H. Pung, "Ontology based context modeling and reasoning using owl," in Pervasive Computing and Communications Workshops, 2004. Proceedings of the Second IEEE Annual Conference on. IEEE, 2004, pp. 18–22

[12] H. Chen, T. Finin, and A. Joshi, "An ontology for context-aware pervasive computing environments," The Knowledge Engineering Review, vol. 18, no. 03, pp. 197–207, 2003.

[13] Georgia M. Kapitsaki, George N. Prezerakos, Nikolaos D. Tselikas, Iakovos S. Venieris "Context-aware service engineering: A survey", The Journal of Systems and Software (2009) 1285–1297.

[14] Kjær, K. E. 2007. A survey of context-aware middleware. In Proc. of the 25th conference on IASTED International Multi-Conference: Software Engineering, 148-155.

[15] A. K. Dey, "Supporting the construction of context-aware applications", In Dagstuhl Seminar on Ubiquitous Computing, Dagsthul, Germany, September 2001

[16] K. Henricksen, J. Indulska, T. McFadden and S. Balasubramaniam, "Middleware for distributed context-aware systems", Proceedings of the International Symposium on Distributed Objects and Applications (DOA), volume 3760 of Lecture Notes in Computer Science, Springer, pp. 846-863, 2005.

[17] Hirschfeld, R., Costanza, P., Nierstrasz, O. Context-oriented programming. Journal of Object Technology, pp 125–151 (March/April), 2008.

[18] E. Tanter, K. Gybels, M. Denker and A. Bergel, "Context-aware aspects", Proceedings of Software Composition 2006, LNCS 4089, Springer Verlag, pp. 227-242.

[19] Q. Z. Sheng and B. Benatallah, "ContextUML: A UML-based modeling language for model-driven development of context-aware web services," Proc. 4th Int. Conf. on Mobile Business, IEEE Comput. Soc., Sydney, Australia, July 2005, pp. 206-212.

[20] D. Ayed, D. Delanote, Y. Berbers, 2007. "MDD approach for the development of context-aware applications," in the 6th Int. and Interdisciplinary Conf. on Modeling and Using Context, Roskilde Univ., Denmark.

[21] C. Simons, "CMP: A UML Context Modeling Profile for Mobile Distributed Systems", Proceedings of the 40 th Hawaii International Conference on System Sciences (HICSS'07), Waikoloa, Big Island, Hawaii, 2007. IEEE Computer Society.

[22] Gu T., Pung H. K., and Zhang D. Q. " A middleware for building contextaware mobile services. " In Proceedings of IEEE Vehicular Technology Conference (VTC), 2004, Milan, Italy.

[23] Dockhorn Costa, Patrícia, Architectural Support for Context-Aware Applications: From Context Models to Services Platforms, CTIT Ph.D.-Thesis Series, No. 07-108, Telematica Instituut Fundamental Research Series, No. 021 (TI/FRS/021), The Netherlands, 2006.

[24] Kapitsaki, G.M., Kateros, D.A., Venieris, I.S., 2008. Architecture for provision of context-aware web applications based on web services. In: Proceedings of IEEE International Symposium on Personal, Indoor and Mobile Radio Communications (PIMRC'08), Cannes, France, pp. 1–5.

[25] Hatim Hafiddi, Hicham Baidouri, Mahmoud Nassar and Abdelaziz Kriouile; 'Context-Awareness for Service Oriented Systems', IJCSI International Journal of Computer Science Issues, Vol. 9, Issue 5, No 2, September 2012 ISSN (Online): 1694-0814 www.IJCSI.org

[26] AKKIRAJU, Rama, FARRELL, Joel, MILLER, John A., et al. Web service semantics-wsdl-s. 2005.

[27] D. Martin, M. Burstein, J. Hobbs, O. Lassila, D. McDermott, S. McIlraith, S. Narayanan, M. Paolucci, B. Parsia, T. Payne et al., "Owl-s: Semantic markup for web services," W3C member submission, vol. 22, pp. 2007–04, 2004.